\documentclass[twocolumn,preprintnumbers,amsmath,amssymb]{revtex4}

\usepackage{graphicx}
\usepackage{dcolumn}
\usepackage{bm}

\begin{document}
\title{Stability of the quantized circulation of an attractive Bose-Einstein condensate in a rotating torus}

\author{Rina Kanamoto}
\author{Hiroki Saito}
\author{Masahito Ueda}
\affiliation{Department of Physics, Tokyo Institute of Technology, Tokyo 152-8551, Japan}
\affiliation{CREST, Japan Science and Technology Corporation (JST), Saitama 332-0012, Japan}

\date{\today}

\begin{abstract}
We investigate rotational properties of a system of bosons with attractive interactions confined in a one-dimensional torus. 
Two kinds of ground states, uniform-density and bright-soliton states, are obtained analytically as functions of the strength of interaction 
and of the rotational frequency of the torus. The quantization of circulation appears in the uniform-density state, 
but disappears upon formation of the soliton. 
By comparison with the results of exact diagonalization of the many-body Hamiltonian, we show that 
the Bogoliubov theory is valid at absolute zero over a wide range of parameters. 
At finite temperature we employ the exact diagonalization method to examine 
how thermal fluctuations smear the plateaus of the quantized circulation. 
Finally, by rotating the system with an axisymmetry-breaking potential, 
we clarify the process in which the quantized circulation becomes thermodynamically stabilized. 
\end{abstract}
\maketitle

\section{Introduction}

Superfluidity represents the assembly of complex phenomena such as persistent currents, 
quantization of circulation, nonclassical rotational inertia, and topological excitations~\cite{Leggett1,Leggett2}. 
Gaseous Bose-Einstein condensates (BECs) offer a testing ground for these phenomena 
because of their great flexibility to realize various experimental conditions. 
For instance, the Feshbach technique makes it possible to control the sign and strength of interactions. 
Furthermore, optical and magnetic traps offer ideal containers of BECs, in which 
microscopic surface rugosities that give rise to dissipation are either absent or can be 
manipulated as tunable parameters~\cite{MITvortex}. 
These experiments have now become possible in low-dimensional systems~\cite{1Dbec1,1Dbec2,2Dbec,bsoliton1,bsoliton2} 
by tightening the confinement in one- or two- direction(s). 
Low-dimensional systems are simple theoretical models for studying vortices, persistent currents, and 
solitons~\cite{Bloch,KPPS,Rokhsar,UL,KPS,Kas1,Kas2,ZS,Carr,SU}. 
For the case of attractive interactions, BECs do not collapse and form solitons in one dimension~\cite{ZS,bsoliton1,bsoliton2}. 
Interestingly, a bright soliton forms also in two dimensions if the strength of interaction is made to oscillate rapidly~\cite{SU}. 

In this paper we study a system of attractive bosons confined in a one-dimensional torus~\cite{toroidal} under rotation. 
When the excitation in the radial direction is negligible, such a system is described by the Lieb-Liniger model~\cite{LL} with 
attractive interaction, in which a rotating term is added to the Hamiltonian and the periodic boundary condition in a finite system is explicitly taken into account. 
When the rotational frequency of the container is increased, a Hartree-Fock theory~\cite{UL} shows that the angular momentum of the system is found to exhibit 
plateaus of the quantized circulation~\cite{HessFairbank,Onsager,Feynman}.  
We investigate the stability of quantized circulation by employing the Gross-Pitaevskii mean-field theory (MFT), 
the Bogoliubov theory, and the exact diagonalization of the many-body Hamiltonian. 
In Refs.~\cite{KSU,Kav}, it is found that quantum fluctuations become significant near the boundary between the uniform and soliton phases 
in a non-rotating torus, and that the boundary is singular in the Bogoliubov approximation. 
We will show, however, that in the rotating torus quantum fluctuations are significant only in the immediate vicinity of a critical point 
at which a normalized rotational frequency of the container $\Omega$ is integral and the dimensionless strength of interaction $\gamma$ is equal to -1/2; 
for other values of $\Omega$ and $\gamma$, there are no singularities in physical quantities at the phase boundary 
because the soliton can be formed without passing through the singular critical point. 

This paper is organized as follows. 
In Sec.~\ref{derivation}, the ground-state wave functions of the system under rotation are derived analytically within the Gross-Pitaevskii MFT. 
In Sec.~\ref{quantum fluctuation}, effects of quantum fluctuations on the quantized circulation are examined based on the Bogoliubov theory 
and on the exact diagonalization of the many-body Hamiltonian. The results obtained by these two methods will be shown to agree very well, 
demonstrating the validity of the Bogoliubov theory. 
In Sec.~\ref{circulation}, the circulation is calculated as a function of $\Omega$ at zero and finite temperatures based on 
the MFT and on the exact diagonalization method. 
In Sec.~\ref{stirring}, the response of the system, which is initially at rest, to a time-dependent axisymmetry-breaking potential is 
examined in order to clarify the process in which the quantized circulation becomes thermodynamically stabilized. 
\nopagebreak
\section{Analytic solution of the one-dimensional Gross-Pitaevskii equation with a rotating drive}\label{derivation}
\subsection{Hamiltonian for the system}\label{formulation}
We consider a system of $N$ identical bosons with mass $M$ which are contained in a rotating torus of radius $R$ 
and cross section $S$, where the angular frequency of rotation is $2\Omega$ in units of $\hbar/2MR^2$.  
The Hamiltonian for the system in the rotating frame of reference is then given by 
\begin{eqnarray}
\hat{{\cal K}}\!=\!\!\!\int_0^{2\pi}\!\!\!\!\!d\theta \!\!\left[\hat{\psi}^{\dagger}(\theta)(\hat{L}-\Omega )^2
\hat{\psi}(\theta)
+\frac{U}{2}\hat{\psi}^{\dagger}(\theta)\hat{\psi}^{\dagger}(\theta)\hat{\psi}(\theta)\hat{\psi}(\theta)\right],\nonumber\\
\label{HamiltonianK}
\end{eqnarray}
where $\hat{L}\equiv -i\partial/\partial\theta$ is the angular-momentum operator, $\theta$ is the azimuthal angle, 
and $U=8\pi aR/S$ characterizes the strength of interaction, where $a$ is the $s$-wave scattering length. 
Here and henceforth, the length, the energy, and the angular momentum are measured in units of $R$, $\hbar^2/2MR^2$, and $\hbar$, respectively. 
In Eq.~(\ref{HamiltonianK}) we include for convenience the kinetic energy $\Omega^2\int_0^{2\pi}d\theta {\hat \psi}^{\dagger}{\hat \psi}$ 
of the rigid body which is a constant and only shifts the zero of energy. 

\begin{figure}
\includegraphics[scale=0.4]{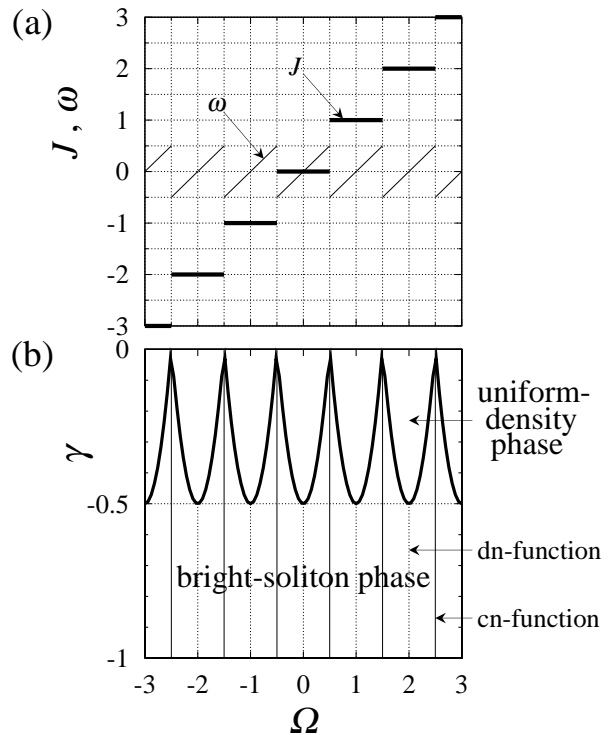}
\caption{(a) Phase winding number $J$ and angular frequency (relative to $J$) $\omega=\Omega-J$ as functions of $\Omega$. 
(b) Ground-state phase diagram. The bold curve corresponds to the phase boundary determined by Eq.~(\ref{stability condition}): 
$\gamma-2(\Omega-J)^2+1/2=0$. 
The upper and lower regions of the bold curve correspond to the uniform-density phase and the bright-soliton phase, respectively. 
The winding number $J$ takes on a constant integer in between the vertical solid lines in Fig.~\ref{pd}(b) and changes by 1 when crossing them. 
On the solid lines, the ground-state wave function is described by the Jacobian cn-function which has one node. 
On the vertical dotted lines, the ground state wave function is described by the Jacobian dn-function. 
In between the horizontal dotted line and the bold curve, the uniform-density state is thermodynamically unstable, and 
below the horizontal dotted line, the uniform-density state is dynamically unstable.}
\label{pd}
\end{figure}

It is shown~\cite{Leggett1,Leggett2}, by using the single-valuedness boundary condition of the many-body wave function 
that physical quantities of the system described by the Hamiltonian~(\ref{HamiltonianK}) change periodically with respect to $\Omega$ with the period of one. 
The phase is therefore characterized by a phase winding number 
\begin{eqnarray}
J=\left[\Omega+\frac{1}{2}\right],
\end{eqnarray}
where the symbol $[x]$ expresses the maximum integer that does not exceed $x$, 
and by a continuous variable 
\begin{eqnarray}\label{omega}
\omega\equiv\Omega-J, 
\end{eqnarray}
which is the angular frequency relative to $J$, and 
the range of $\omega$ is limited to $-1/2 \le \omega < 1/2$. 
In Fig.~\ref{pd}(a), we show $J$ and $\omega$ as functions of $\Omega$. 


We first seek the lowest-energy state of the one-dimensional Gross-Pitaevskii equation (GPE) in the rotating frame of reference, 
\begin{eqnarray}\label{GPeq}
\left[(\hat{L}-\Omega)^2
+2\pi\gamma|\psi(\theta)|^2\right]\psi(\theta)=\mu\psi(\theta), 
\end{eqnarray}
where $\mu$ is the chemical potential, and a dimensionless parameter 
\begin{eqnarray}
\gamma\equiv \frac{UN}{2\pi},
 \end{eqnarray}
gives the ratio of the mean-field interaction energy to the zero-point kinetic energy. 
The condensate wave function $\psi(\theta)$ is assumed to obey the periodic boundary condition $\psi(0)=\psi(2\pi)$, 
and is normalized as $\int_0^{2\pi}|\psi(\theta)|^2d\theta=1$.
It is appropriate to assume the form of the solution as 
\begin{eqnarray}\label{general solution form}
\psi(\theta)=\sqrt{\rho(\theta)}\ e^{i\varphi(\theta)}, 
\end{eqnarray}
where $\rho$ is the number density and $\varphi$ the phase. 

\subsection{Uniform-density solution}\label{uniform-density solution}

The stationary state which circulates on the ring with a uniform density is a plane wave, 
\begin{eqnarray}\label{plane wave}
\psi(\theta)=\sqrt{\frac{1}{2\pi}}\ e^{iJ\theta},\qquad\mu=\omega^2+\gamma. 
\end{eqnarray}

The stability of the ground state is determined by the sign of the lowest excitation energy. 
Diagonalizing the Bogoliubov-de Gennes (BdG) equations 
\begin{eqnarray}
\left[\left(-i\frac{\partial}{\partial\theta}-\Omega\right)^2-\mu+4\pi\gamma|\psi|^2\right]u_n+2\pi\gamma\psi^2 v_n&=&
\!\!\lambda_n u_n, \nonumber\\
\left[\left(i\frac{\partial}{\partial\theta}-\Omega\right)^2-\mu
+4\pi\gamma|\psi|^2\right]v_n+2\pi\gamma\psi^{\ast 2} u_n&=&\!\!-\lambda_n v_n, \nonumber\\
\label{BdG}
\end{eqnarray}
we obtain the excitation energies $\lambda_n$ and the corresponding amplitudes $u_n,v_n$ with positive norms as 
\begin{eqnarray}
\lambda_{n}&=&\sqrt{n^2(n^2+2\gamma)}-2n\omega\label{lambdaP},\\
u_n&=&{\cal N}^+_n e^{i(J+n)\theta},\\
v_n&=&{\cal N}^-_n e^{-i(J-n)\theta},
\end{eqnarray}
where the normalization constants ${\cal N}_n^{\pm}$ are determined from the orthonormality condition 
\begin{eqnarray}
\int_0^{2\pi}[u_n(\theta)u_m^*(\theta)-v_n(\theta)v_m^*(\theta)]d\theta=\delta_{n,m}, 
\end{eqnarray}
as 
\begin{eqnarray}
{\cal N}_n^{\pm}=\sqrt{\frac{1}{4\pi}\left[\frac{n^2+\gamma}{\sqrt{n^2(n^2+2\gamma)}}\pm 1\right]}.
\end{eqnarray}

For repulsive interactions $\gamma >0$, the ground state of Eq.~(\ref{GPeq}) always takes the form of the uniform-density solution~(\ref{plane wave}), 
since the lowest excitation energy $\lambda_{-1}$ (for $-1/2 \le \omega <0$) or 
$\lambda_{1}$ (for $0 \le \omega < 1/2$) from it is positive for all $\gamma$ and $\Omega$. 
However, for the attractive case, the first excitation energy $\lambda_{1}$ or $\lambda_{-1}$ becomes zero at 
\begin{eqnarray}\label{stability condition}
\gamma-2\omega^2+\frac{1}{2}=0. 
\end{eqnarray}
In Fig.~\ref{pd}(b), we show the phase diagram of the ground states with respect to $\gamma$ and $\Omega$. 
The phase boundary~(\ref{stability condition}) of the ground states is represented by the bold curve. 
The uniform-density solution becomes thermodynamically unstable for 
$-1/2 \le \gamma < 2\omega^2-1/2$ (in between the horizontal dotted line and the bold curve) because the first excitation energy 
becomes negative, while it becomes dynamically unstable for $\gamma < -1/2$ (in the lower region of the horizontal dotted line) 
because the first excitation energy acquires an imaginary part. 
The uniform-density state is thus stable only when $\gamma \ge 2\omega^2-1/2$ (i.e., in the upper region of the bold curve). 

\subsection{Bright-soliton solution}\label{soliton}

For $\gamma < 2\omega^2-1/2$ (in the lower region of the bold curve in Fig.~\ref{pd}(b)), the uniform-density state is either thermodynamically 
or dynamically unstable, and a soliton state becomes stable. 
In this subsection, we derive the ground-state soliton solution of Eq.~(\ref{GPeq}). 
The definitions of several kinds of elliptic integrals and elliptic functions~\cite{math} used throughout this paper are summarized in appendix~\ref{elliptic} 
to make the present paper self-contained. 

Substituting Eq.~(\ref{general solution form}) into the GPE~(\ref{GPeq}) and taking the real and the imaginary parts, we obtain
\begin{eqnarray}
\mu=\Omega^2-\frac{(\sqrt{\rho})''}{\sqrt{\rho}}+(\varphi')^2-2\Omega\varphi'+2\pi \gamma \rho,\label{1st_eq}\\
\varphi''+2\varphi '\frac{(\sqrt{\rho})'}{\sqrt{\rho}}-2\Omega\frac{(\sqrt{\rho})'}{\sqrt{\rho}}=0\label{2nd_eq}. 
\end{eqnarray}
Equation~(\ref{2nd_eq}) is integrated to give 
\begin{eqnarray}
\varphi'=\Omega+\frac{W}{\rho}\label{phaseD}. 
\end{eqnarray}
Substituting this into Eq.~(\ref{1st_eq}) yields
\begin{eqnarray}\label{eq of mu}
\mu\rho^2=\pi \gamma\rho^3+V\rho-\left(\frac{\rho'}{2}\right)^2-W^2. 
\end{eqnarray}
This equation can be rewritten in the form of an elliptic integral 
\begin{eqnarray}
\int d\theta=\int \frac{d\rho}{\sqrt{4\pi \gamma\rho^3-4\mu\rho^2+4V\rho-4W^2}} 
\end{eqnarray}
which has formally the same solution as the case without rotating term~\cite{Carr} and is given by 
\begin{eqnarray}
\rho(\theta)={\cal N}^2
\left[{\rm dn}^2\left(\left.\frac{K}{\pi}(\theta-\theta_0)\right|m\right)-\eta m_1\right],\label{g_density}
\end{eqnarray}
where dn$(u|m)$ is the Jacobian elliptic function, $0 \le \eta \le 1$, and $m_1=1-m$. 
A constant $\eta$ is given below in Eq.~(\ref{eta}) and the parameter $m$ will be determined later in Eq.~(\ref{mgeneral}). 
We denote the complete elliptic integrals of the first and the second kinds as $K\equiv K(m)$ and $E\equiv E(m)$, respectively. 
Since the soliton breaks the translational symmetry, the solution~(\ref{g_density}) contains an arbitrary parameter $\theta_0$ 
which is regarded as the center of mass of the total particles. 
From the normalization condition, the normalization constant ${\cal N}$ is determined as ${\cal N}^2=K/\left[2\pi (E-\eta m_1 K)\right]$.  
Substituting $\rho$ in Eq.~(\ref{g_density}) in Eq.~(\ref{eq of mu}), we obtain 
\begin{eqnarray}
{\cal N}^2&=&\frac{K}{2\pi (E-\eta m_1 K)}=\frac{K^2}{\pi^3 |\gamma|},\\
\mu&=&\frac{1}{2\pi^2}(f_{\rm d}-f_{\rm c}-f)\label{gCP},\\
\eta&=&\frac{f_{\rm d}}{2m_1K^2}=1-\frac{f_{\rm c}}{2m_1K^2}\label{eta},\\
W&=&-{\rm sgn}(\omega)\sqrt{\frac{ff_{\rm c}f_{\rm d}}{8\pi^8\gamma^2}}\label{WW}, 
\end{eqnarray}
where $f,f_{\rm c},f_{\rm d}$ are defined as
\begin{eqnarray}
f&\equiv&2K^2-2KE-\pi^2\gamma,\label{f}\\
f_{\rm c}&\equiv&2m_1K^2-2KE-\pi^2\gamma,\label{fcn} \\
f_{\rm d}&\equiv&2KE+\pi^2\gamma,\label{fdn}
\end{eqnarray}
and sgn denotes the sign function such that ${\rm sgn}(\omega)=-1$ when $-1/2 < \omega < 0$ (i.e., $J-1/2 < \Omega < J$), 
and ${\rm sgn}(\omega)=+1$ when $0 < \omega < 1/2$, (i.e., $J < \Omega <J+1/2$, see Fig.~\ref{pd}(a)). 
The cases of $\omega=0,\pm 1/2$ will be treated separately below. 

From Eqs.~(\ref{phaseD}) and (\ref{g_density}), the phase part is given by
\begin{eqnarray}
\varphi (\theta)=\int d\tilde{\theta}\qquad\qquad\qquad\qquad\qquad\qquad\qquad\qquad\qquad\nonumber\\
\!\!\!\!\!\left[\Omega+\!
\left\{\frac{K^2}{\pi^3 |\gamma|W}\left[{\rm dn}^2
\left(\left.\frac{K}{\pi}\tilde{\theta}\right| m\right)-\eta m_1\right]\right\}^{-1}\right],\quad
\end{eqnarray}
which is integrated analytically using the relations (\ref{Jrelations}) and (\ref{Pi1}) to give 
\begin{eqnarray}\label{g_phase}
\varphi(\theta)=\Omega\theta-{\rm sgn}(\omega)\delta_2^{-1}\Pi\!\left(\left.n;\frac{K\theta}{\pi}\right| m\right),
\end{eqnarray}
where $\Pi(n;u|m)$ is the elliptic integral of the third kind, and the constants $n$ and $\delta_2$ are given by
\begin{eqnarray}
n=\frac{m}{1-\eta m_1}=\frac{2mK^2}{f}, \qquad
\delta_2=K\sqrt{\frac{2f}{f_{\rm c}f_{\rm d}}}.
\end{eqnarray}
Since the wave function is single-valued, the phase $\varphi$ satisfies 
\begin{eqnarray}\label{svbc}
\varphi(2\pi)-\varphi(0)=2\pi J. 
\end{eqnarray}
This condition can be used to determine the value of $m$ as follows. 
In the present case, the parameters $m$ and $n$ satisfy the relation $m < n <1$, and then using the relations~ (\ref{reductionPi})-(\ref{lambda function}), 
the elliptic integral $\Pi(n;u|m)$ at $\theta=2\pi$ reduces to
\begin{eqnarray}
\Pi(n;2K|m)
&=&2\left[K+\frac{\pi}{2}\delta_2\left\{1-\Lambda_0(\varepsilon |m)\right\}\right],\\
\varepsilon&=&\arcsin \sqrt{\frac{f_{\rm c}}{m_1 f}},
\end{eqnarray}
where $\Lambda_0$ is Heuman's Lambda function defined in terms of 
the incomplete elliptic integrals of the first $F(\varepsilon|m)$ and the second kinds $E(\varepsilon|m)$ as 
\begin{eqnarray}
\Lambda_0(\varepsilon |m)&=&\frac{2}{\pi}\left[KE(\varepsilon|m_1)-(K-E)F(\varepsilon|m_1)\right].
\end{eqnarray}
The phase at $\theta=2\pi$ thus becomes
\begin{eqnarray}\label{phase2pi}
\varphi(2\pi)=2\pi\Omega-{\rm sgn}(\omega)\left[\sqrt{\frac{2f_{\rm d}f_{\rm c}}{f}}+\pi(1-\Lambda_0)\right].
\end{eqnarray}
Using the notation $\omega\equiv \Omega-J$ instead of $\Omega$ and $J$, condition~(\ref{svbc}) then leads to  
\begin{eqnarray}\label{mgeneral}
2\pi |\omega|=\sqrt{\frac{2f_{\rm d}f_{\rm c}}{f}}+\pi(1-\Lambda_0), 
\end{eqnarray}
which determines the parameter $m$. This equation has a solution only when $\gamma < 2\omega^2-1/2$ and $0< |\omega | <1/2$, 
i.e., in the region delimited by adjacent vertical dotted and solid lines and by the bold curve in Fig.~\ref{pd}(a). 

Next let us consider two special cases: $\omega= 0$ 
($\Omega$ is equal to integer $J$, i.e., on the vertical dotted lines in Fig.~\ref{pd}(b)) and 
$|\omega| = 1/2$ ($\Omega$ is equal to half-integer $J+1/2$, i.e., on the vertical solid lines). 
The ground-state wave function given by Eqs.~(\ref{g_density}) and (\ref{g_phase}) is simplified when $|\omega|= 0$ and $|\omega|= 1/2$ 
according to the limiting values of several parameters summarized in appendix~\ref{limit}. 

When $\omega=0$, the amplitude $\sqrt{\rho(\theta)}$ reduces to the Jacobian elliptic function ${\rm dn}(u|m)$, 
and the phase $\varphi (\theta)$ reduces to $J\theta$, 
\begin{eqnarray}
\psi(\theta)=\sqrt{\frac{K^2}{\pi^3|\gamma|}}\ 
{\rm dn}\!\left(\left. \frac{K(\theta-\theta_0)}{\pi}\right|m\right)e^{iJ\theta}. \label{dng}
\end{eqnarray}
Substituting this solution into the GPE~(\ref{GPeq}) yields the chemical potential 
\begin{eqnarray}
\mu=-\frac{K^2}{\pi^2}(1+m_1),\label{dnCP}
\end{eqnarray}
and an equation 
\begin{eqnarray}
f_{\rm d}=2KE+\pi^2\gamma=0\label{mdn}, 
\end{eqnarray}
which determines the parameter $m$ only when $\gamma < -1/2$. 
The dn-solution includes the ground state in the rest container with $\Omega=0$. 

When $|\omega|=1/2$, the amplitude is given by the Jacobian elliptic function cn$(u|m)$, and $\varphi (\theta)$ 
reduces to $(J+1/2)\theta$,
\begin{eqnarray}\label{cng}
\psi(\theta)=\sqrt{\frac{K^2m}{\pi^3|\gamma|}}\ \ 
\left|{\rm cn}\!\left(\left.\frac{K(\theta-\theta_0)}{\pi}\right|m\right)\right|e^{i\left(J+\frac{1}{2}\right)\theta}.
\end{eqnarray}
The chemical potential and an equation that determines the parameter $m$ 
are obtained as 
\begin{eqnarray}
\mu&=&-\frac{K^2}{\pi^2}(1-2m_1)\label{cnCP},\\
f_{\rm c} &=& 2m_1K^2-2KE-\pi^2 \gamma=0. \label{mcn}
\end{eqnarray}
The equation~(\ref{mcn}) has a solution only when $\gamma < 0$. 
The wave function~(\ref{cng}) has a node at $\theta=\theta_0+\pi$ and the phase jumps by an amount of $\pi$ at the node. 
With increasing $\Omega$ adiabatically, the vortex enters the ring through this node. 

\subsection{Ground-state properties}\label{properties}

\begin{figure}
\includegraphics[scale=0.45]{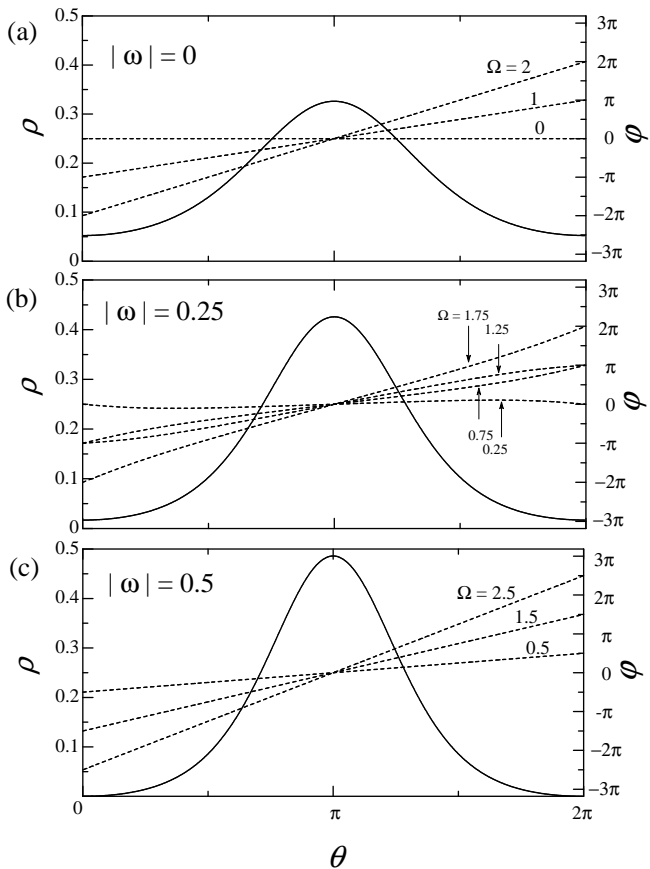}
\caption{Densities $\rho$ (solid curves) and phases $\varphi $ (dotted curves) for 
several values of angular frequency $\omega$ and phase winding number $J=\Omega-\omega$ with $\gamma=-0.55$. 
The density depends only on $|\omega|$. The phase difference $\varphi (\theta +2\pi)-\varphi (\theta)$ is given by $2\pi J$ in Figs.~(a) and (b), 
and by $2\pi J+\pi$ in Fig.~(c) because at $|\omega|=0.5$ the wave function has a node at which the phase jumps by an amount of $\pi$.}\label{WFfig}
\end{figure}

\begin{figure}
\includegraphics[scale=0.39]{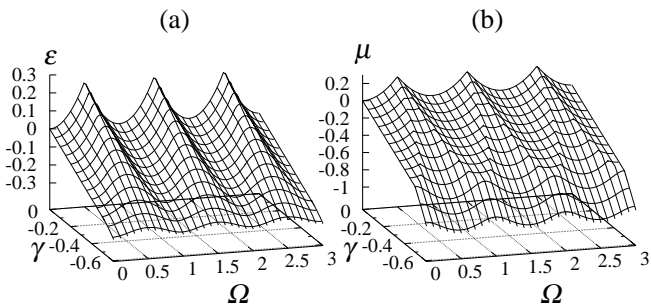}
\caption{(a) Ground-state energy ${\cal E}$ per atom, and (b) chemical potential $\mu$. 
Both of them decrease monotonically with increasing $|\gamma|$ and are periodic with respect to $\Omega$.}
\label{EneCP}
\end{figure}

To illustrate the $\Omega$ dependence of the soliton solution, we present in Fig.~\ref{WFfig} the densities and the phases for $\gamma=-0.55$, 
where the integral constant of the phase is chosen so that $\varphi(\pi)=0$. 
The density profiles of the solitons depend only on the relative angular frequency $|\omega|$. 
As the strength of interaction is increased, the parameter $m$ approaches unity 
(see Fig.~\ref{m} in appendix~\ref{limit}) for all $\omega$. 
In the limit of $m\to 1$, both dn$(u|m)$ and cn$(u|m)$ become sech$u$ 
which is a soliton solution in infinite space~\cite{ZS}. 

The ground-state energy ${\cal E}$ per atom of the uniform-density state is 
\begin{eqnarray}
{\cal E}=\omega^2+\frac{\gamma}{2}, 
\end{eqnarray}
and that of the soliton is 
\begin{eqnarray}
{\cal E}&=&\gamma+\frac{K\left[3E-(1+m_1)K\right]}{\pi^2}\nonumber\\
&\ &+\frac{2K^2\left[3E^2-2(1+m_1)KE+m_1K^2\right]}{3\pi^4\gamma}.\label{gE}
\end{eqnarray}
Equation~(\ref{gE}) reduces to the energy of the dn-solution in the limit $\omega\to 0$ as 
\begin{eqnarray}
{\cal E}=-\frac{K^2\left[(1+m_1)E+m_1K\right]}{3\pi^2 E},\label{dnE}
\end{eqnarray}
and to that of cn-solution in the limit $|\omega|\to 1/2$ as
\begin{eqnarray}
{\cal E}=-\frac{K^2\left[(1-2m_1)E-m_1(2-3m_1)K\right]}{3\pi^2(E-m_1K)}.\label{cnE}
\end{eqnarray}
The ground-state energy per atom ${\cal E}$ and the chemical potential $\mu$ are shown in Figs.~\ref{EneCP}(a) and (b). 
For a given $\gamma$, the ground-state energy ${\cal E}$ reaches minima for integer $\Omega$ and maxima for half-integer $\Omega$, 
and is smooth everywhere. 
In the regime $\gamma < -1/2$, the chemical potential $\mu$ becomes maximal for integer $\Omega$ and minimal 
for half-integer $\Omega$, and has kinks at the phase boundaries given by $\gamma-2(\Omega-J)^2+1/2=0$. 
The type of the phase transition is the same as that of the non-rotating case~\cite{KSU}; at the phase boundary, 
(i) ${\cal E}$ is smooth, the first derivative of ${\cal E}$ with respect to $\gamma$ or $\Omega$ has a kink, and the second derivative 
of ${\cal E}$ has a jump, and 
(ii) $\mu$ has a kink, and the first derivative of $\mu$ has a jump. 
Detailed behaviors of these quantities near the phase boundary are discussed in appendix~\ref{phase_boundary}.

\section{Effects of quantum fluctuations}\label{quantum fluctuation}

Without rotation, the quantum depletion diverges at $\gamma=-1/2$ and 
the first excitation energy obtained by the Bogoliubov theory becomes gapless at that point, 
which requires a modification of the MFT~\cite{KSU,Kav}. 
We investigate here whether in the presence of rotation there is such a singular point 
at which the effects beyond the Bogoliubov theory are significant.

\begin{figure} [b]
\includegraphics[scale=0.45]{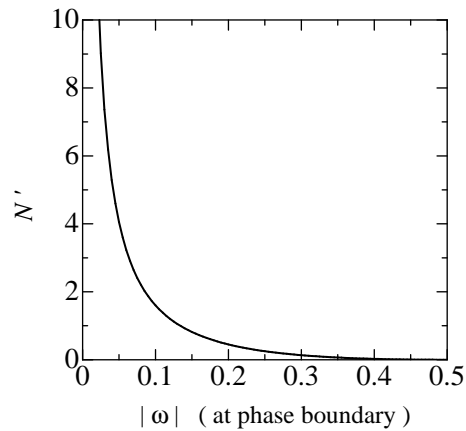} 
\caption{Quantum depletion $N'$ (Eq.~(\ref{depletion})) at the phase boundary calculated by the Bogoliubov theory.} 
\label{depletion figure} 
\end{figure} 

We evaluate the depletion of the condensate calculated by the Bogoliubov theory as 
\begin{eqnarray}
N'=\int_0^{2\pi}\sum_{n\ne 0}|v_n(\theta)|^2 d\theta, 
\end{eqnarray}
where $v_n(\theta)$ is the hole amplitude in the BdG equation~(\ref{BdG}). 
If $N'/N$ is of the order of unity, the validity of the Bogoliubov theory is not ensured. 
Since the excitation in the uniform-density regime is contributed mainly from the excitation with quantum number 1, 
the depletion in the uniform-density regime becomes
\begin{eqnarray}
N'\simeq \frac{1+\gamma}{\sqrt{1+2\gamma}}-1. 
\end{eqnarray}
At the phase boundary $\gamma=2\omega^2-1/2$, this is rewritten as 
\begin{eqnarray}\label{depletion}
N'\simeq \omega+\frac{1}{4\omega}-1, 
\end{eqnarray}
which is shown in Fig.~\ref{depletion figure}. 
The quantum depletion diverges at the boundary $\omega=0,\gamma=-1/2$. 
However, as $|\omega|$ is increased, $N'$ at the phase boundary decreases and the depletion becomes much less pronounced as shown in 
Fig.~\ref{depletion figure}. 
This result is also inferred from Fig.~\ref{pd}(b). 
The uniform-density state becomes dynamically unstable below the horizontal dotted line $\gamma=-1/2$, 
and this line touches the phase boundary (the bold curve) only at $\omega=0$. 

\begin{figure} [b]
\includegraphics[scale=0.5]{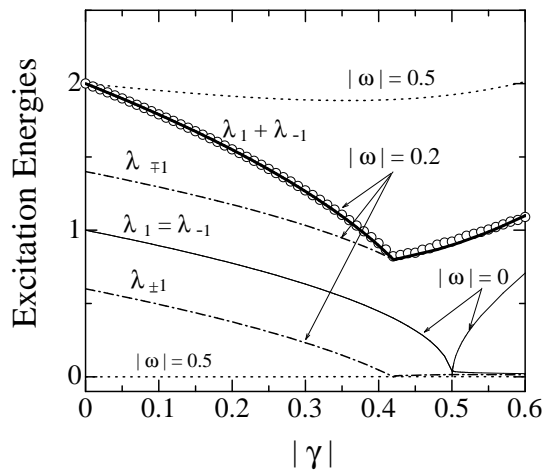} 
\caption{Bogoliubov spectra $\lambda_n$ for $|\omega|=0$ (solid curve), $|\omega|=0.2$ (dashed-and-dotted curves), and $|\omega|=0.5$ (dotted curves). 
The (quasi) zero-energy levels in the soliton regime correspond to the Goldstone modes associated with the translation-symmetry breaking 
due to soliton formation. 
Open circles show the excitation energy ${\cal E}_1({\cal L}_0)-{\cal E}_0({\cal L}_0)$ obtained by the 
exact diagonalization of the Hamiltonian for $N=500$ with $|\omega|=0.2$, which agrees with the Bogoliubov spectrum 
$\lambda_1+\lambda_{-1}$ (bold curve) even at the phase boundary $|\gamma|=0.42$. } 
\label{spectra} 
\end{figure} 

We compare low-lying energy levels obtained by the Bogoliubov theory 
with those obtained by the exact diagonalization of the many-body Hamiltonian~(\ref{HamiltonianK}). 
In Fig.~\ref{spectra}, we show the Bogoliubov spectra $\lambda_n$ obtained by the BdG equation~(\ref{BdG}) 
as a function of the strength of interaction for the angular frequency $|\omega|=0,0.2$, and $0.5$. 
The curves represent the excitation energies from the uniform-density state for $|\gamma| < -2\omega^2+1/2$, and 
those from the soliton for $|\gamma| > -2\omega^2+1/2$. 
All levels are continuous at the boundaries $|\gamma| =-2\omega^2+1/2$, which indicates a smooth crossover between the uniform-density 
state and the soliton state. When $\omega=0$ (solid curves), $\lambda_1$ and $\lambda_{-1}$ are degenerate 
in the uniform-density regime, but they separate into two branches, the first excitation and 
a Goldstone mode, in the soliton regime ($|\gamma| > 0.5$). 
The Goldstone mode, which is associated with the translation-symmetry breaking, boosts the soliton along the ring without increasing the energy. 
As $|\omega|$ is increased from zero (dashed-and-dotted curves), the degeneracy is lifted by rotation. 

Next we calculate the low-lying energy levels by the exact diagonalization of the many-body Hamiltonian to see 
how they deviate from the Bogoliubov spectra near the phase boundary. 
The procedure of the diagonalization is the same as that without rotation~\cite{KSU}. 
Denoting the number of atoms with angular momentum $k$ as $n_k$, the plane-wave bases are prepared as 
$|n_{l_0-l_{\rm c}},\dots, n_{l_0-1},n_{l_0},n_{l_0+1},\dots,n_{l_0+l_{\rm c}}\rangle$ where 
$l_0\simeq J$ is the angular momentum of the condensate, and $l_{\rm c}$ is the cutoff. 
Within the subspace in which the particle number and the total angular momentum are conserved as 
\begin{eqnarray}
\sum_{k=l_0-l_{\rm c}}^{l_0+l_{\rm c}} n_k=N,\qquad \sum_{k=l_0-l_{\rm c}}^{l_0+l_{\rm c}} kn_k={\cal L}, 
\end{eqnarray}
we perform the diagonalization of the Hamiltonian~(\ref{HamiltonianK}), which is rewritten as
\begin{eqnarray}
\hat{{\cal K}}=\sum_{l}(l-\Omega)^2\hat{c}_l^{\dagger}\hat{c}_l+\frac{\gamma}{2N}\sum_{klmn}\hat{c}^{\dagger}_k
\hat{c}_l^{\dagger}\hat{c}_m\hat{c}_n\delta_{m+n-k-l}\label{HamiltonianD}. 
\end{eqnarray}
We denote the ground-state angular momentum as ${\cal L}_0$ that gives the lowest energy eigenvalue, and the ground-state energy as ${\cal E}_0({\cal L}_0)$. 
Open circles in Fig.~\ref{spectra} show excitation energy ${\cal E}_1({\cal L}_0)-{\cal E}_0({\cal L}_0)$ 
obtained by the diagonalization of the Hamiltonian~(\ref{HamiltonianD}) for $N=500$ and $|\omega|=0.2$ with cutoff $l_c=1$. 
The results agree very well with the Bogoliubov spectrum $\lambda_1+\lambda_{-1}$ 
(we represent here only the excitation that conserve the total angular momentum ${\cal L}_0$). 
We have also confirmed that the difference between the Bogoliubov spectrum and the exact one becomes even 
smaller as $|\omega|$ is further away from zero, consistent with 
the analysis of the depletion of the condensate.
The Bogoliubov theory is thus vindicated except for $\gamma\simeq-1/2$ and $\Omega\simeq J$. 

\section{Quantized circulation at zero and finite temperature}\label{circulation}
\subsection{Quantized circulation at zero temperature}\label{absolute zero}

The magnetic flux is excluded from a superconductor when the applied magnetic field is below a critical value. 
The analog of this Meisner effect in superfluid system is the Hess-Fairbank effect~\cite{HessFairbank}, in which 
the system is not set into rotation when the frequency $\Omega$ of a rotating drive is below a critical value $\Omega_{\rm c}$. 
When $\Omega$ exceeds $\Omega_{\rm c}$, the circulation of the system, which is defined as the integral of the superfluid velocity 
along a closed contour, is quantized in units of $h/M$~\cite{Onsager,Feynman}, 
in analogy with the case of a type-II superconductor in which quantized vortices enter the system when the external magnetic field exceeds the lower critical field 
$H_{\rm c_1}$. 
The applied magnetic field and the magnetic flux in the superconductor correspond to 
the applied rotation and the angular momentum $\langle\hat{L}\rangle$ of the superfluid, respectively. 

\begin{figure}
\includegraphics[scale=0.6]{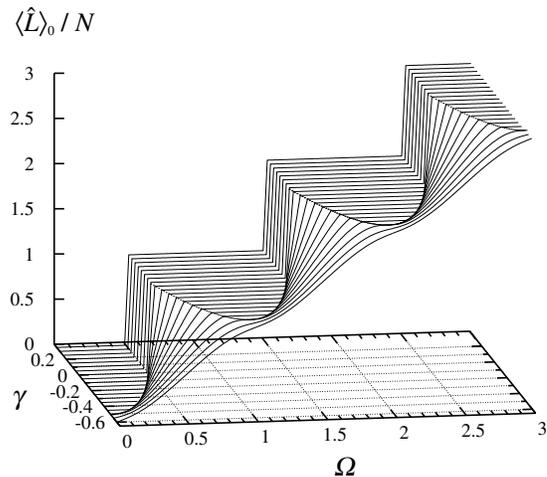}
\caption{Expectation value of the angular momentum per atom at zero temperature. 
The plateau regions correspond to $\gamma \ge 2(\Omega-J)^2-1/2$. 
When $\gamma\le -0.5$, the plateaus disappear.}
\label{L0}
\end{figure}

We calculate the angular momentum of the ground state 
$\langle\hat{L}\rangle _0=N \int_0^{2\pi}\psi^{\dagger}(\theta) \left(-i\partial_\theta\right)  \psi (\theta)d\theta$, 
where $\psi$ is the mean-field solution~(\ref{plane wave}) with Eqs.~ (\ref{g_density}) and (\ref{g_phase}) obtained in Sec.~\ref{derivation}, 
and we find the analytical expression for the expectation value of $\hat{L}$ per atom as
\begin{eqnarray}
\langle \hat{L}\rangle _0/N=
\left\{
\begin{array}{lll}
J,\qquad &\ &\gamma \ge 2\omega^2-1/2,\\
J+\omega+2\pi W,\qquad &\ &\gamma < 2\omega^2-1/2,\\
\end{array}
\right.
\label{mfL}
\end{eqnarray}
where $W$ is given in Eq.~(\ref{WW}) [see also Fig.~\ref{parameters}(a)]. 
In the limits $|\omega|\to 0$ and $|\omega|\to 1/2$, $J+\omega+2\pi W$ reduces to $J$ and $J+1/2$, respectively. 
Equation~(\ref{mfL}) is shown in Fig.~\ref{L0} as a function of $\Omega$ and $\gamma$, where 
the plateaus correspond to the uniform-density regime, and the crossover regions between plateaus correspond to 
the soliton regime. For $\gamma < -1/2$, the stable uniform-density state does not exist and the plateaus disappear. 

Since the quantum fluctuation is small, Eq.~(\ref{mfL}) correctly describes the ground-state angular momentum 
except for $\gamma\simeq -1/2$ and $\Omega\simeq J$. 
Both Hartree-Fock theory~\cite{UL} and Monte Carlo calculation~\cite{Kar} 
show that this system exhibits the Hess-Fairbank effect in a certain parameter regime, 
which is consistent with our results. 
 
We briefly comment on the same problem for repulsive interaction ($\gamma > 0$). 
According to the MFT, the ground state with repulsive interaction has a uniform density for all parameters. 
The expectation value of the angular momentum then increases stepwise like the noninteracting case with $\gamma=0$ in Fig.~\ref{L0}. 
The repulsive bosons do not prefer the mixing of different angular-momentum states since it costs the Fock exchange energy. 

\subsection{Quantized circulation at finite temperature}\label{finiteT}

We examine the effect of thermal fluctuations on the quantized circulation at finite temperature.
The total angular momentum of the system is obtained from the derivative of free energy ${\cal F}$ with respect to $\Omega$ as 
\begin{eqnarray}\label{L at T}
\langle {\hat L} \rangle _{\tau}=-\frac{1}{2}\frac{\partial{\cal F}}{\partial\Omega}+N\Omega, 
\end{eqnarray}
where
\begin{eqnarray}\label{tau}
\tau=\frac{k_{\rm B}T}{\hbar^2/(2MR^2)}, 
\end{eqnarray}
is a dimensionless temperature with $k_{B}$ being the Boltzmann constant, 
$\langle \cdots \rangle _{\tau}$ denotes the ensemble average at temperature $\tau$, and 
the second term in Eq.~(\ref{L at T}) corresponds to the angular momentum of the rigid body arising 
from the constant term of the Hamiltonian~(\ref{HamiltonianK}). 

\begin{figure}
\includegraphics[scale=0.47]{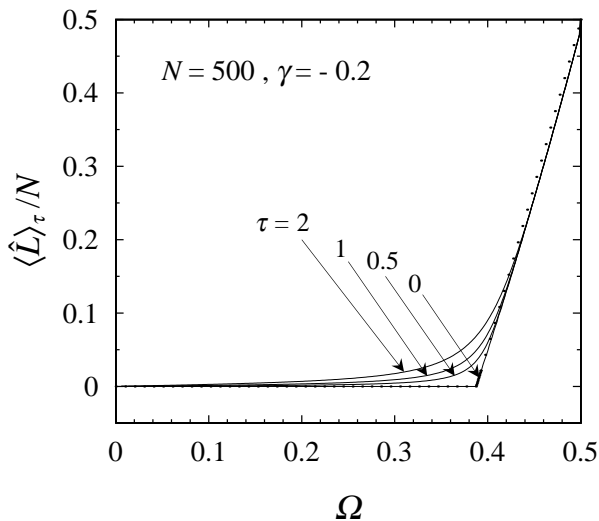}
\caption{Angular momentum per atom for $\gamma=-0.2$ and $N=500$ at temperatures 
$\tau=0,0.5,1,$ and $2$ obtained by diagonalization of Hamiltonian~(\ref{HamiltonianD}). 
Dotted lines show the result of the mean-field theory at $\tau=0$. } 
\label{LT}
\end{figure}

To evaluate the free energy, we employ the exact diagonalization method of the many-body Hamiltonian. 
All low-lying levels ${\cal E}_n$ have been obtained in Sec.~\ref{quantum fluctuation}, and we use them to calculate the free energy 
\begin{eqnarray}\label{FE}
{\cal F}=-\tau \ln \sum_{n} e^{-{\cal E}_n/\tau}. 
\end{eqnarray}
Figure~\ref{LT} shows the angular momentum for several temperatures calculated from Eqs.~ (\ref{L at T}) and (\ref{FE}), 
where the mean-field result at $\tau=0$ is also presented for comparison as dotted lines. 
At absolute zero, the result obtained by the exact diagonalization agrees well with the mean-field result, 
which supports the validity of the MFT consistently with the results in Sec.~\ref{quantum fluctuation}. 
As the temperature is increased, thermal excitations wash out the edge of 
the step of the circulation, and the angular momentum of the system approaches 
that of the classical fluid, i.e., $\langle \hat{L} \rangle _{\tau}=N\Omega$. 

In the present paper, we have focused on the properties near $\tau=0$ in this calculation; however, in principle, this method can be extended to 
higher temperature by increasing the cutoff angular momentum as long as the excitations of radial modes are negligible. 
Near $|\omega|\simeq 1/2$, though the diagonalization gives less accurate results than at $\omega\simeq 0$ because a larger number of 
bases are needed in order to allow the state to have a node, the accuracy can also be improved by the increase of the cutoff angular momentum. 

Finally, we consider an experimental situation. To be concrete, let us consider $^{7}$Li; then Eq.~(\ref{tau}) leads to 
\begin{eqnarray}
T\simeq 34\frac{\tau}{(R\  [\mu{\rm m}])^2}\ [{\rm nK}]. 
\end{eqnarray}
Thus for a torus with $R=1\ \mu{\rm m}$, $\tau=2$ in Fig.~{\ref{LT}} corresponds to $T=68$ nK which can be achieved with current experimental techniques.
A torus geometry may also be set up by the technique of microelectronic chips~\cite{MP}. 

\section{Preparation of the ground state by a stirring potential}\label{stirring}

\begin{figure}
\includegraphics[scale=0.45]{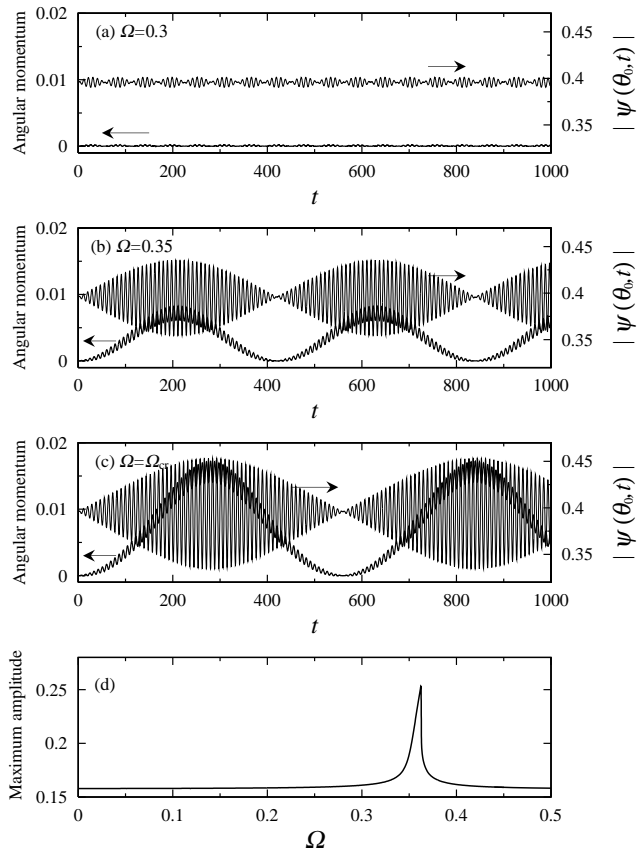}
\caption{Time evolutions of the angular momentum of the condensate (bold curves) and of the amplitude $|\psi(\theta_0,t)|$ at $\theta_0=\pi$ (solid curves) 
driven by a time-dependent stirring potential~(\ref{potential}) with $\gamma=-1/4$ for 
(a) $\Omega=0.3$, (b) $\Omega=0.35$, and (c) $\Omega=\Omega_{\rm cr}$. 
Note that the oscillations of the density and angular momentum are enhanced close to $\Omega_{\rm cr}=\sqrt{1/8}\simeq 0.354$.
(d) Maximum amplitude of the wave function as a function of $\Omega$.} 
\label{resonance}
\end{figure}

Rotation of the system can actually be driven by a potential that breaks the axisymmetry of the system. 
As a concrete example, we consider a time-dependent potential 
\begin{eqnarray}\label{potential}
V(\theta,t)=V_0\cos(\theta-2\Omega t), 
\end{eqnarray}
which stirs the system with angular frequency $2\Omega$, and we fix the amplitude of the potential as $V_0=10^{-3}$. 
We take as an initial state $\psi(\theta,t=0)=1/\sqrt{2\pi}$ with a fixed strength of interaction $\gamma=-1/4$, 
and let the system evolve in time according to the GPE, 
\begin{eqnarray}
i\frac{\partial}{\partial t}\psi(\theta,t)=\hspace{5cm}\nonumber\\
\left[-\frac{\partial^2}{\partial \theta^2}+V(\theta,t)+2\pi\gamma|\psi(\theta,t)|^2\right]\psi(\theta,t). 
\end{eqnarray}
Figures~\ref{resonance} (a)-(c) show the time evolutions of the angular momentum of the condensate and the amplitude $|\psi(\theta=\pi,t)|$. 
One of the phase boundaries for $\gamma=-1/4$ determined by Eq.~(\ref{stability condition}) is $\Omega_{\rm cr}=\sqrt{1/8}\simeq 0.354$. 
As $\Omega$ gets close to $\Omega_{\rm cr}$, 
the stirring potential causes the dynamical instability, and a significant growth in density is seen in Fig.~\ref{resonance}(c). 
Figure ~\ref{resonance}(d) plots the maximum amplitude $|\psi(\theta_0)|$ as a function of $\Omega$, 
which shows the resonance induced near the phase boundary. 
However, the angular momentum of the system oscillates without damping, and the system does not reach any stationary state. 

\begin{figure}
\includegraphics[scale=0.43]{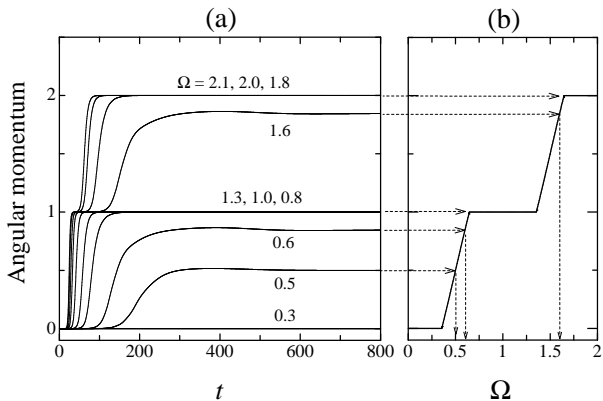}
\caption{(a) Time evolution of the angular momentum of an initially uniform state with a damping constant 
$\Gamma=0.1$ for $\gamma=-1/4$ and $V_0=10^{-3}$. 
(b) Expectation value of angular momentum per atom obtained by the MFT for $\gamma=-1/4$.}
\label{vortex}
\end{figure}

To achieve a stationary state, we must introduce energy dissipation. We therefore study the time-evolution of the system 
according to a generalized GPE~\cite{CMB,TKU}, 
\begin{eqnarray}
(i-\Gamma)\frac{\partial}{\partial t}\psi(\theta,t)=\hspace{5cm}\nonumber\\
\left[\left(i\frac{\partial}{\partial\theta}+\Omega\right)^2\!\!+V_0 \cos\theta+2\pi\gamma|\psi(\theta,t)|^2\right]\!\!\psi(\theta,t), 
\end{eqnarray}
where $\Gamma$ is a phenomenological damping constant. 
Figure~\ref{vortex} (a) shows the time evolutions of the angular momentum for several values of the stirring angular frequency with $\Gamma=0.1$. 
The system acquires a finite angular momentum, and 
the magnitude of the absorbed angular momentum converges to that of the thermodynamically stable state as shown in Fig.~\ref{vortex}(b). 
We note that as long as $\Omega$ lies in the same plateau region, 
the angular momentum converges to the same integral value, demonstrating that 
the ground state derived in Sec.~\ref{derivation} can indeed be prepared by the time-dependent stirring potential in the presence of energy dissipation 
and that the circulations is indeed quantized in the thermodynamically stable state. 

From these results, it is concluded that the system reaches a thermodynamically stable state after both density fluctuations and energy dissipation. 
The oscillations in the density is drastically enhanced near the resonant frequency (dynamical instability). 
However, the system can acquire a net angular momentum due to energy dissipation (Landau instability). 
Similar mechanisms were found in the vortex lattice formation~\cite{Kas2}. 

\section{Conclusions}

In conclusion, we investigated the rotational properties of one-dimensional bosons with attractive interactions confined in a rotating torus trap. 

We derived the ground state wave function analytically within the MFT as a function of the strength of interaction $\gamma$ 
and of the rotational frequency of the torus $\Omega$. 
A uniform-density solution and a bright-soliton one are found to smoothly cross over to each other. 
The density of the soliton depends on the rotational frequency $\omega$ relative to $J$ and has a node at $\Omega=J+1/2$ when a vortex enters the ring. 

In order to investigate the validity of the Bogoliubov theory, we compared the excitation spectrum obtained by the Bogoliubov theory with 
that obtained by the exact diagonalization of the many-body Hamiltonian, 
and found that the MFT well describes the ground state except for $\Omega\simeq J, \gamma\simeq -1/2$, 
where the phase boundary and the borderline of the onset of the dynamical instability coincide, and hence quantum fluctuations are significant. 

The angular momentum of the ground state is quantized with respect to $\Omega$ in the uniform-density regime, but it is not in the soliton-regime. 
The circulations at finite temperature are examined by the exact diagonalization method. 

To understand the process in which the system is set into rotation, we considered a time-dependent stirring potential which breaks the axisymmetry of the torus. 
The potential is shown to induce the dynamical instability, causing the density to oscillate resonantly and triggering the acquisition of the angular momentum 
by the system. With energy dissipation, the quantized circulation is found to become thermodynamically stabilized via the Landau instability. 

\begin{acknowledgments}
This research was supported by a Grant-in-Aid for Scientific Research (Grant No.15340129), by Special 
Coordination Funds for Promoting Science and Technology from the Ministry of Education, 
Culture, Sports, Science and Technology of Japan and by the Yamada Science Foundation. 
\end{acknowledgments}

\appendix
\section{Elliptic integrals and elliptic functions}\label{elliptic}

We summarize definitions and interrelations of some elliptic integrals and elliptic functions~\cite{math}
that are used in the present paper. 

Elliptic integrals of the first kind $F$, the second kind $E$, and the third kind $\Pi$ are defined as
\begin{eqnarray}
F(\phi\backslash\alpha)\!\!&\equiv&\!\!\int _0^{\phi}\frac{d \theta}{\sqrt{1-\sin ^2\alpha \sin^2\theta}},\label{F2}\\
E(\phi\backslash\alpha)\!\!&\equiv&\!\!\int_0^{\phi}d\theta\sqrt{1-\sin^2\alpha\sin^2\theta},\label{E2}\\
\Pi(n;\phi\backslash\alpha)\!\!&\equiv&\!\!\int_0^\phi\!\frac{d\theta}{(1-n\sin^2\theta)\sqrt{1-\sin^2\alpha\sin^2\theta}}\label{Pi2}.
\end{eqnarray}
Using a parameter $m\equiv\sin^2\alpha$, Jacobian elliptic functions are defined by
\begin{eqnarray}\label{Jdefinition}
{\rm sn}(u|m)&\equiv&\sin \phi,\\
{\rm cn}(u|m)&\equiv&\cos \phi,\\
{\rm dn}(u|m)&\equiv&\sqrt{1-m\sin ^2 \phi}, 
\end{eqnarray}
and these functions are interrelated as
\begin{eqnarray}\label{Jrelations}
{\rm dn}^2 (u|m)-m_1=m{\rm cn}^2 (u|m)=m\left[1-{\rm sn}^2(u|m)\right].
\end{eqnarray}
Using the parameter set $\{u,m\}$ instead of $\{\phi,\alpha\}$, the elliptic integrals are also expressed as 
\begin{eqnarray}
F(u|m)&\equiv&\int _0^u dv=u,\label{F1}\\
E(u|m)&\equiv&\int _0^u {\rm dn}^2 (v|m) dv,\label{E1}\\
\Pi(n;u|m)&\equiv&\int_0^u\frac{dv}{1-n{\rm sn}^2(v|m)}\label{Pi1}.
\end{eqnarray}

Elliptic integrals are said to be complete when $\phi=\pi/2$, and are usually denoted as 
\begin{eqnarray}
K(m)\equiv F\left(\left.\frac{\pi}{2}\right|m\right), 
\qquad E(m)\equiv E\left(\left.\frac{\pi}{2}\right|m\right)\label{complete}.
\end{eqnarray}
Complete elliptic integrals of the first and second kinds are expanded for $|m| <1 $ as infinite series 
\begin{eqnarray}
K(m)\!\!\!&=&\!\!\!\frac{\pi}{2}\left[1\!+\!\!\left(\frac{1}{2}\right)^{\!2}\!\!m\right.
\!+\!\!\left(\frac{1\!\cdot \!3}{2\!\cdot \!4}\right)^{\!2}\!\!m^2
\left.\!+\!\!\left(\frac{1\!\cdot \!3\!\cdot\! 5}{2 \!\cdot\! 4\! \cdot\! 6}\right)^{\!2}\!\!m^3\!+\!\cdots\right],\label{series1}\nonumber\\
\\
E(m)\!\!\!&=&\!\!\!\frac{\pi}{2}\left[1\!-\!\!\left(\frac{1}{2}\right)^{\!2}\!\!m\right.
\!-\!\!\left(\frac{1\!\cdot\! 3}{2\!\cdot\! 4}\right)^{\!2}\!\!\frac{m^2}{3}
\left.\!-\!\!\left(\frac{1\!\cdot\! 3\!\cdot\! 5}{2\!\cdot\! 4\! \cdot\! 6}\right)^{\!2}\!\!\frac{m^3}{5}\!-\!\cdots\right].\label{series2}\nonumber\\
\end{eqnarray}
The elliptic integral of the third kind $\Pi(n;\phi\backslash\alpha)$ has other expressions depending on the relation between $m$ and $n$. 
When $m < n <1$ and $\phi =\pi/2$, it reduces to 
\begin{eqnarray}
\Pi(n;\frac{\pi}{2}\backslash\alpha)
&=&K(\alpha)+\frac{\pi}{2}\delta_2\left\{1-\Lambda_0(\varepsilon\backslash\alpha)\right\}\label{reductionPi},\\
\delta_2&\equiv&\sqrt{\frac{n}{(1-n)(n-\sin^2 \alpha)}}\label{delta2},\\
\varepsilon&\equiv&\arcsin \sqrt{\frac{1-n}{\cos ^2\alpha}}\label{varepsilon},
\end{eqnarray}
where $\Lambda_0$ is Heuman's lambda function defined as 
\begin{eqnarray}\label{lambda function}
\Lambda_0(\phi\backslash\alpha)&\equiv&\frac{2}{\pi}\left[K(\alpha)E(\phi\backslash 90^{\circ}-\alpha)\right.\nonumber\\
&-&\left.\left\{K(\alpha)-E(\alpha)\right\}F(\phi\backslash 90^{\circ}-\alpha)\right].
\end{eqnarray}

\section{Limiting behaviors of the soliton solutions}\label{limit}

We consider the limits $|\omega|\to 0$ and $|\omega| \to 1/2$ of the soliton solution for $0 <|\omega| <1/2$ given by 
\begin{eqnarray}
\rho(\theta)&=&\frac{K^2}{\pi^3|\gamma|}
\left[{\rm dn}^2\left(\left.\frac{K}{\pi}(\theta-\theta_0)\right|m\right)-\eta m_1\right]\label{densityA},\\
\varphi(\theta)&=&\Omega\theta-{\rm sgn}(\omega)\delta_2^{-1}\Pi \left(\left. n;\frac{K\theta}{\pi}\right| m\right)\label{phaseA},\\
\mu&=&\frac{1}{2\pi^2}(f_{\rm d}-f_{\rm c}-f)\label{CPA}, 
\end{eqnarray}
where 
\begin{eqnarray}
f&\equiv&2K^2-2KE-\pi^2\gamma,\label{fA}\\
f_{\rm c}&\equiv&2m_1K^2-2KE-\pi^2\gamma,\label{fcnA} \\
f_{\rm d}&\equiv&2KE+\pi^2\gamma.\label{fdnA}
\end{eqnarray}

\begin{figure}
\includegraphics[scale=0.43]{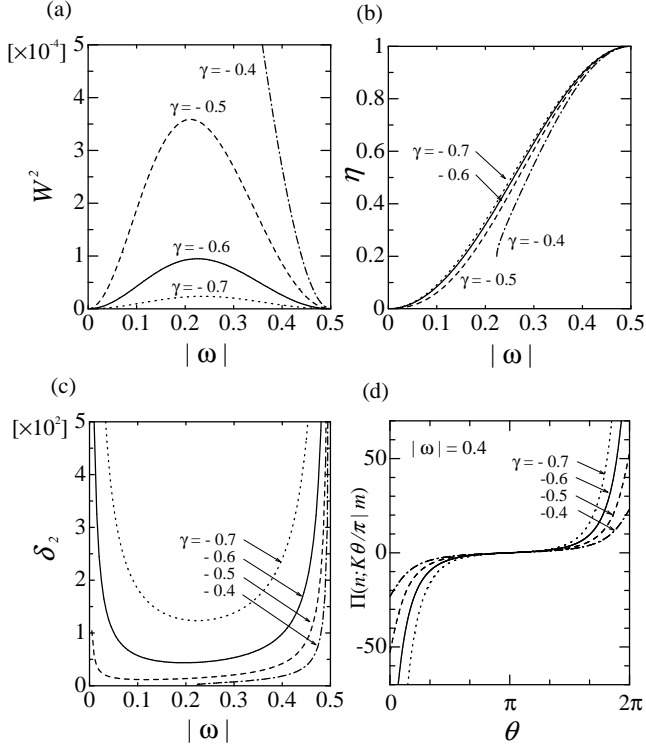}
\caption{(a) (b) (c) Behaviors of several parameters appeared in the soliton solution for $\gamma=-0.4$ (dashed-and-dotted), 
$\gamma=-0.5$ (dashed), $\gamma=-0.6$ (solid), and $\gamma=-0.7$ (dotted curves). 
(d) Incomplete elliptic integral of the third kind $\Pi(n;K\theta/\pi |m)$ for $|\omega|=0.4$ as a function of $\theta$.}
\label{parameters}
\end{figure}

The limiting values and behaviors of several parameters are summarized in Table~\ref{limit table} and Fig.~\ref{parameters}, respectively. 
In the limit $|\omega|\to 0$ ($f_{\rm d}\to 0$), the equations~(\ref{densityA}) and (\ref{phaseA}) continuously become those of the dn-solution 
since $\eta\to 0$ and $W\to 0$ from Fig.~\ref{parameters} and Table~\ref{limit table}. 
In the limit $|\omega|\to 1/2$ ($f_{\rm c}\to 0$), where the parameters behave as 
$\eta \to 1$, $W\to 0$, Eqs.~(\ref{densityA}) and (\ref{phaseA}) reproduce the cn-solution. 
It can easily be verified by setting $f_{\rm d}=0$ or $f_{\rm c}=0$ in Eq.~(\ref{CPA}) that in these limits 
the chemical potential for $0<|\omega|<1/2$ continuously approaches 
\begin{eqnarray}
\mu\!\!&=&\!\!-\frac{K^2}{\pi^2}(1+m_1)=\frac{1}{2\pi^2}(-f_{\rm c}-f),\ \  |\omega|=0, \label{dnCPA}\\
\mu\!\!&=&\!\!-\frac{K^2}{\pi^2}(1-2m_1)=\frac{1}{2\pi^2}(f_{\rm d}-f),\ \  |\omega|=1/2.\label{cnCPA}
\end{eqnarray}

The first derivative of the ground-state energy with respect to $\Omega$ has a kink at the phase boundary, 
which is verified by the relations~(\ref{mfL}) and (\ref{L at T}) as, 
\begin{eqnarray}\label{derE}
\frac{\partial{\cal E}}{\partial\Omega}&=&2\left(\Omega-\langle {\hat L} \rangle _0 /N\right)\nonumber\\&=&
\left\{
\begin{array}{lll}
\!\!\displaystyle{0},\qquad &\ &|\omega|=0, 1/2,\\
\!\!\displaystyle{-4\pi W},\quad&\ &0 < |\omega| < 1/2.
\end{array}
\right.
\end{eqnarray}

\begin{table}
\begin{tabular}{cccc}\hline\hline
\ & \ \ \ $|\omega|\to 0$\ \ \ &\ \ \  $|\omega|\to 0.5$\ \ \  &\ \ \  $m\to$ 0\ \ \ \\
\hline
\ \ \ \ \ $\eta$\ \ \ \ \  & 0 & 1 & $1+2\gamma$\\
\ \ \ \ \ $n$\ \ \ \ \ & $m$ & 1 & 0\\
\ \ \ \ \ $W^2$\ \ \ \ \ & 0 & 0 & $\sqrt{1+2\gamma}/4\pi$\\
\ \ \ \ \ $\delta_2$\ \ \ \ \ & $\infty$ & $\infty$ & $\sqrt{1/(1+2\gamma)}$\\
\ \ \ \ \ $\varepsilon$\ \ \ \ \ & $\pi/2$ & 0 & $\pi/2$\\
\hline\hline
\end{tabular}
	\caption{Limiting values of various parameters of the soliton solution. 
	The limit $|\omega|\to 0$ is equivalent to the limit $f_{\rm dn}\to 0$ or 
	$\gamma\to -2KE/\pi^2-0$, and $|\omega|\to 0.5$ is equivalent to $f_{\rm cn}\to 0$ or $\gamma\to -2(KE-m_1K^2)/\pi^2-0$. 
	The limit $m\to 0$ corresponds to the uniform-density limit $|\gamma|\to -2\omega^2+1/2+0$.}
	\label{limit table}
\end{table}

\begin{figure}
\includegraphics[scale=0.45]{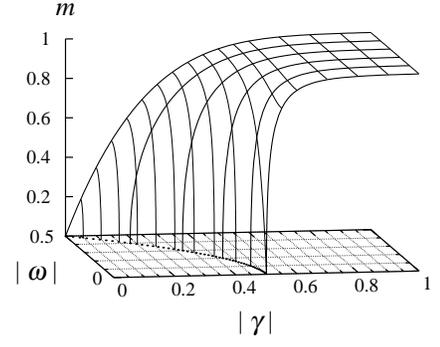}
\caption{The solution $m$ calculated numerically from~Eq.~(\ref{m eqs}). 
The dotted curve on the $\omega$-$\gamma$ plane represents the phase boundary $\gamma-2\omega^2+1/2=0$, 
and there exists a unique solution $m$ in the soliton regime $|\gamma| >- 2\omega^2+1/2$.}
\label{m}
\end{figure}

Figure~\ref{m} shows the solution $m$ calculated numerically by solving 
\begin{eqnarray}\label{m eqs}
\left.\qquad
\begin{array}{lll}
\displaystyle{f_{\rm d} =0},&\ &|\omega|=0,\\
\displaystyle{2\pi |\omega| \!=\!\! \sqrt{\frac{2f_{\rm d}f_{\rm c}}{f}}\!+\!\pi (1-\Lambda_0)},&\ &\!\!\!0 < |\omega| < 1/2,\qquad\\
\displaystyle{f_{\rm c} =0},&\ &|\omega|=1/2, 
\end{array}
\right.
\end{eqnarray}
where the line of intersection between the curves and the $\gamma$-$\omega$ plane 
corresponds to the phase boundary $\gamma-2\omega^2+1/2=0$.

\section{Ground-state properties near the phase boundary}\label{phase_boundary}

We investigate the continuity of the ground-state properties at the phase boundary.  
From Fig.~\ref{m}, we see that the parameter $m$ becomes zero for all soliton solutions in the limit 
$|\gamma| \to -2\omega^2+1/2+0$. 
The uniform-density limit of the soliton solution is hence mathematically obtained by taking the limit of $m\to 0$. 
Let $\delta$ be a positive small deviation of $\gamma$ from the value at the phase boundary for a fixed $\omega$. 
For $|\omega|=0,1/2$, the parameter $m,K(m), E(m)$, and hence the physical quantities near the phase boundary are expanded in terms of $\delta$ as follows.  

For the dn-solution ($\omega=0$), the phase boundary is $\gamma=-1/2$ and let $\gamma=-1/2-\delta$. 
Using the equation $f_{\rm d}=0$, we obtain
\begin{eqnarray}
m=8\delta^{1/2}-32\delta+89\delta^{3/2}-200\delta^2+O(\delta^{5/2})\label{dnM}. 
\end{eqnarray}
The expansion formulae~(\ref{series1}) and (\ref{series2}) become 
\begin{eqnarray}
K\!\!&=&\!\!\frac{\pi}{2}\!\left(1+2\delta^{1/2}+\delta+\frac{1}{4}\delta^{3/2}
+\frac{1}{2}\delta^2\right)\!+O(\delta^{5/2})\label{dnK},\\
E\!\!&=&\!\!\frac{\pi}{2}\!\left(1-2\delta^{1/2}+5\delta-\frac{33}{4}\delta^{3/2}
+\frac{23}{2}\delta^2\right)\!+O(\delta^{5/2}).\nonumber\\
\label{dnE}
\end{eqnarray}
Using these expansions, the ground-state energy and the chemical potential are 
expressed as functions of $\delta$ instead of $m$ near the phase boundary $\gamma=-1/2-\delta$. 
The jump of the second derivative of the ground-state energy and the first derivative of the chemical potential are obtained as 
\begin{eqnarray}
{\cal E}''_{\rm dn}-{\cal E}''_{\rm u}=-4,\qquad \mu'_{\rm dn}-\mu'_{\rm u}=-2, 
\end{eqnarray}
at $\gamma=-1/2$ and $\omega=0$. 

For the cn-solution ($|\omega|=1/2$) at $\gamma=-\delta$, the expansions are obtained in a similar way as
\begin{eqnarray}
m&=&4\delta-6\delta^2+O(\delta^3)\label{cnM},\\
K&=&\frac{\pi}{2}\left(1+\delta+\frac{3}{4}\delta^2\right)+O(\delta^3)\label{cnK},\\
E&=&\frac{\pi}{2}\left(1-\delta+\frac{3}{4}\delta^2\right)+O(\delta^3)\label{cnE}.
\end{eqnarray}
The ground-state energy and the chemical potential are expanded in terms of $\delta$ as
\begin{eqnarray}
{\cal E}_{\rm cn}&=&\frac{1}{4}-\frac{3\delta}{4}-\frac{\delta^2}{8}+O(\delta^3),\\
\mu_{\rm cn}&=&\frac{1}{4}-\frac{3\delta}{2}-\frac{3\delta^2}{8}+O(\delta^3)
\end{eqnarray}
which are continuously connected with those of the uniform-density solution at $\gamma=0$ and $|\omega|=1/2$. 
However, the first derivative of the ground-state energy and that of the chemical potential are given by 
\begin{eqnarray}
{\cal E}'_{\rm cn}&=&-\frac{3}{4}-\frac{\delta}{4}+O(\delta^2), \\
\mu'_{\rm cn}&=&-\frac{3}{2}-\frac{3\delta}{4}+O(\delta^2), 
\end{eqnarray}
and have discontinuous jumps at $\gamma=0$ and $|\omega|=1/2$ by the following amounts: 
\begin{eqnarray}
{\cal E}'_{\rm cn}-{\cal E}'_{\rm u}=-\frac{1}{4},\qquad\mu'_{\rm cn}-\mu'_{\rm u}=-\frac{1}{2}.
\end{eqnarray}


\end{document}